# Two distinct charge density wave orders and emergent superconductivity in pressurized CuTe

Shuyang Wang,[1,2,5] Qing Wang,[1,2,5] Chao An,[4] Yonghui Zhou,[1,2,3] Ying Zhou,[4] Xuliang Chen (陈绪亮),[1,2,3,6,*] Ning Hao,[1,2,3,*] and Zhaorong Yang[1,2,3,4,*]

[1]Anhui Province Key Laboratory of Condensed Matter Physics at Extreme Conditions, High Magnetic Field Laboratory, HFIPS, Chinese Academy of Sciences, Hefei 230031, China

[2]Science Island Branch of Graduate School, University of Science and Technology of China, Hefei 230026, China

[3]Collaborative Innovation Center of Advanced Microstructures, Nanjing 210093, China

[4]Institutes of Physical Science and Information Technology, Anhui University, Hefei 230601, China

[5]These authors contributed equally

[6]Lead Contact

*Correspondence**:** xlchen@hmfl.ac.cn (X.C.); haon@hmfl.ac.cn (N.H.); zryang@issp.ac.cn (Z.Y.)

SUMMARY

The discovery of multiple charge-density-wave (CDW) orders in superconducting cuprates and Kagome $CsV_3Sb_5$ has offered a unique milieu for studying the interplay of CDW and superconductivity and altered our perspective on their nature. Here, we report a high-pressure study of quasi-one-dimensional CDW material CuTe through ultralow-temperature (400 mK) electrical transport and temperature-dependent Raman spectroscopy measurements and first-principles calculations. We provide solid evidence that the pristine CDW order (CDW1) transforms into a distinct CDW order (CDW2) at ~6.5 GPa. Calculations show that the driving force of CDW1 is due to the nesting effect and that of CDW2 probably arises from the electronic correlated interaction. Strikingly, pressure-induced superconductivity is observed with a dome-like phase diagram and its transition displays an extraordinary broadening along with the crossover from CDW1 to CDW2. These results demonstrate that pressurized CuTe provides a promising playground for understanding the intricated interplay of multiple CDWs and superconductivity.

**INTRODUCTION**

Understanding the interplay between superconductivity and other collective electronic phenomena has always been one of the central issues within the condensed-matter and material-physics communities.[1-5] An example is the interplay between superconductivity and the charge-density-wave (CDW) order in low-dimensional materials.[6,7] Intuitively, superconductivity should be favored by suppressing the CDW because the closure of the CDW energy gap contributes more charge carriers and increases the density of states at the Fermi surface within the Bardeen-Cooper-Schrieffer scenario.[8,9] However, in real materials, the relationship between superconductivity and CDW is very complicated, involving coexistence, competition, and cooperation under external tuning parameters like doping, intercalation, or pressure.[7,10-24] For example, $TiSe_2$ and $TaS_2$ are both prototypical CDW transition-metal dichalcogenides. With the application of pressure, superconductivity is found to emerge near the CDW quantum critical point in the former case[11] while it seems to be independent of the CDW order in the latter case.[12] In a recently discovered kagome superconductor, $CsV_3Sb_5$, a double-dome-like superconducting phase diagram is observed with unusual interplay between superconductivity and CDW under high pressure, and moreover a new dome-like superconducting phase reemerges after the vanishing of the CDW upon further compression.[13-19]

CuTe, as a two-dimensional layered material, has received much recent attention[25-31] because it displays a quasi-one-dimensional CDW order below ~346 K. In the high-temperature non-CDW state, CuTe has a *Pmmn* (SG. 59) structure, consisting of wrinkled Cu planes sandwiched in an array of quasi-one-dimensional Te chains. Upon cooling below the CDW transition temperature $T_{CDW}$, a periodic lattice modulation occurs within the quasi-one-dimensional Te chains.[25,26] At the same time, a CDW energy gap opens up due to the well-nested quasi-one-dimensional Fermi surface sheets contributed by $p_x$ orbitals of Te.[26,30] Associated with formation of the CDW order, the Raman spectra of CuTe exhibit one collective amplitude mode and four zone-folded (ZF) modes in addition to two phonon modes.[30] Our recent high-pressure investigation of CuTe showed that the application of pressure can effectively suppress the CDW order.[31] Moreover, at a high pressure far above the critical pressure where the CDW disappears, superconductivity is induced due to a structural phase transition.[31] Interestingly, the preliminary electrical transport (1.8-300 K) in the intermediate pressure region reveals subtle modification of the CDW order—the resistivity anomaly related to the CDW transition changes its shape at a characteristic pressure.[31]

Here, we provide solid evidence by combined electrical transport and Raman spectroscopy measurements that the pristine CDW order (CDW1) transforms to a distinct CDW order (CDW2) at ca. 6.5 GPa. Furthermore, by extending the temperature down to sub-kelvin, we found pressure-induced superconductivity at ca. 4.8 GPa along with suppression of the CDW1, and the superconducting transition displays anomalous broadening below 10 GPa relating to the crossover from CDW1 to CDW2. According to the calculations, the birth of CDW2 accompanies the emergence of four small isotropic hole-like Fermi pockets with nearly full weight of $p_z$ orbital of Te atoms. In comparison with several different mechanisms for CDW orders, we argue that the CDW2 could be driven by the electronic correlated interaction.

## RESULTS AND DISCUSSION

### Transport evidence for pressure-induced new CDW and superconductivity

Figure 1A shows the temperature dependence of resistivity ($\rho$) for a CuTe single crystal with temperatures down to 400 mK under various pressures up to 13.8 GPa. In the low-pressure range, the original CDW transition leads to a hump-like anomaly at $T_{CDW}$ in the $\rho(T)$ curve and shows as a dip in the corresponding derivative $d\rho/dT$ curve (see Figure 1B). Here, $T_{CDW}$ is defined as the onset temperature of anomalies observed in $d\rho/dT$ upon cooling. In agreement with our previous work,[31] Figure 1A uncovers an unusual pressure-dependent evolution of the CDW order. The resistivity hump gets suppressed gradually with increasing pressure up to 6.7 GPa, indicating suppression of the CDW under pressure. However, as the pressure is increased to 8.1 GPa, one can see that the anomaly in $d\rho/dT$ changes its profile from the original dip to a peak. This observation signals a subtle modification of the CDW order as in the case of pressurized $CsV_3Sb_5$.[13-15] In the pressurized $CsV_3Sb_5$, the $T_{CDW}$ decreases monotonically with increasing pressure across the crossover region,[13-15] while in the pressurized CuTe the $T_{CDW}$ jumps to a higher temperature at the critical pressure. The peak-like anomaly in d$\rho$/d$T$ shifts to lower temperatures upon further compression and is scarcely observed beyond 9.5 GPa. These unusual evolutions of the CDW are captured in other independent runs, as shown in Figure S1, indicating the reproducibility of the results. We also measured Hall resistivity at 5 K and various pressures (see Figure S2). At 0.8 GPa, the Hall resistivity shows quasi-linear field dependence with positive slope, which indicates that hole-type carriers dominate the transport behavior, consistent with the investigations of CuTe at ambient pressure.[28,30] With increasing pressure, the slope of $\rho_{yx}(H)$ first increases and then decreases followed by a sign change between 6.7 and 8.1 GPa.

In addition, by lowering temperature down to 400 mK, we demonstrate that superconductivity is induced, accompanied by the suppression of the pristine CDW order. Figure. 1C shows an enlarged view of the $\rho(T)$ data below 3 K, highlighting a non-monotonic evolution of the superconducting transition with increasing pressure. A very small drop in resistivity starts to appear at 4.8 GPa with an onset temperature of $T_C^{onset}$ ~0.5 K; zero resistance is observed at 5.7 GPa. This pressure-induced superconducting transition is also confirmed by the temperature dependence of resistivity under external magnetic fields at 5.7 GPa (Figure S3), from which the upper critical field at zero temperature is estimated to be 1260 Oe. However, in the pressure range of 5.7-9.5 GPa, the superconducting transition presents an anomalous broadening with $\Delta T_C = T_C^{onset} - T_C^{zero}$ of ~2 K. These observations are confirmed in other runs, as shown in Figure S4. The broadened superconducting transition is also reminiscent of that in pressurized CsV$_3$Sb$_5$, in which the anomalous behavior is attributed to strong competition between superconductivity and the newly appeared CDW order.[13-15] As discussed above, CuTe also exhibits a signature of CDW transformation under pressure. By coincidence, the superconducting transition in pressurized CuTe becomes sharp when the peak-like anomaly in $T$ associated with the CDW transition is hard to discern.

**Raman spectroscopy measurements under high pressure and low temperature**
To further study the evolution of CDW in CuTe, we performed high-pressure Raman experiments at temperatures between 70 and 320 K, as shown in Figure 2. At ambient pressure, there are one collective amplitude mode and four ZF modes which are associated with the CDW order.[30] The amplitude mode is a soft-phonon mode coupled to the electronic density at the CDW wavevector and dressed by the amplitude fluctuations of the CDW order parameter, which is a fingerprint of the CDW order.[32-34] The ZF modes correspond to normal phonons folded to the center of the Brillouin zone. At 1.5 GPa and 70 K, the Raman spectrum consists of six peaks (see Figure 2A). According to Ref. [30], these peaks can be assigned to two phonon modes, $A_g^1$ (130.8 cm$^{-1}$) and $A_g^2$ (142.7 cm$^{-1}$), one CDW amplitude mode (52.7 cm$^{-1}$), and two ZF modes (74.2 and 123.9 cm$^{-1}$). As expected, the amplitude mode softens rapidly upon warming while the ZF modes move at a much slower rate, as shown in Figure 2C. The temperature-dependent frequency of the amplitude mode can be well fitted by a modified mean-field (MMF) model,[30,35] from which $T_{CDW}$ is estimated to be 285 K at 1.5 GPa (Figure 2J).

With increasing pressure, as shown in Figures 2C-2F, the CDW amplitude mode at 70 K shifts gradually to lower frequencies and is undetectable at 6.5 GPa; at the same time, its intensity becomes steadily weaker. $T_{CDW}$, deduced from fits to the MMF model of the temperature evolution of the amplitude mode at each pressure, decreases monotonically under pressure, indicative of suppression of the pristine CDW order (Figures 2J-2L). At 6.5 GPa, the spectrum contains only two $A_g$ modes in the whole measured temperature range (Figure 2F). It is interesting to note that the suppression of the pristine CDW order up to 6.5 GPa seen in the Raman spectra is consistent with that reflected in the electronic transport measurements. And more strikingly, as the pressure is increased to 8.5 GPa, two sets of peaks with CDW character reappear, which is in line with the above conjecture of a CDW transformation. One set of peaks displays a rapid softening with increasing temperature, as shown in Figures 2B and 2G, characteristic of the CDW amplitude mode. The temperature dependence of peak frequency for the emergent CDW can also be fitted by the MMF model, which gives a $T_{CDW}$ of 134 K at 8.5 GPa (Figure 2M). The other set can be assigned to the ZF mode, because its peak position is insensitive to temperature from 70 to 140 K. With further increase of the pressure beyond 9.5 GPa, both the CDW amplitude mode and the ZF mode disappear (Figure 2I), also in agreement with the vanishing of the resistivity anomaly (Figure 1).

One may be concerned whether the soft-mode behavior observed at 8.5 and 9.5 GPa can be of other different origins. The most straightforward one could be a structural transition. We have collected the temperature-dependent Raman spectra at various pressures up to 22.0 GPa (across the $Pmmn\rightarrow Cm$ structural transition occurring at ~20 GPa[31]) and further plotted the calculated and experimental Raman shift versus pressure for the leading modes observed. We can see that the $A_g^1$ and $A_g^2$ modes at 320 K and 70 K increase linearly in the pressure range of 0-16 GPa with no discontinuity and their trends change evidently across the structural transition at 22.0 GPa (see Figures S5 and S6 and Supplemental Note 1 for details). In addition, we measured the pressure dependent $\rho$-$T$ via the thermal cycling between 1.8-300 K (Figure S7) and found no evident temperature hysteretic behavior, irrespective of types of the $\rho$-$T$ anomaly. These two data sets from the Raman and resistivity measurements together suggest that the possibility of a structural transition can be safely ruled out. Alternatively, the soft mode behavior could be related to a ferroelectric transition, such as in $PbTiO_3$ and $BaTiO_3$.[36] Nevertheless, the ferroelectric transition usually refers to a polar structural distortion and metallic ferroelectricity is extremely rare in nature with the only exception of $LiOsO_3$[37,38] and $Sr_{1-x}Ca_xTiO_{3-\delta}$ ($0.002<x<0.009$, $\delta<0.001$)[39], to the best of our knowledge. The Jahn-Teller effect might also lead to a soft-mode behavior as observed in $DyVO_4$ and $RbCoF_3$.[36] This should not be the present case of CuTe as the $Cu^{2+}$ ion is magnetically inactive and no degenerate electronic state is expected.

**Analysis of the complex relations between CDW and superconductivity**

Based on the data from the above comprehensive high-pressure characterizations, we plotted a temperature-pressure phase diagram of CuTe, as shown in Figure 3A. The most distinct feature is the occurrence of a CDW transition from the pristine CDW order (CDW1) to a distinct CDW order (CDW2) at ca. 6.5 GPa, which is consistently revealed by both the electronic transport and Raman measurements. The transition temperatures $T_{CDW1}$ and $T_{CDW2}$ deduced from two methods show similar pressure dependence—i.e., $T_{CDW1}$ decreasing linearly with compression followed by a jump to $T_{CDW2}$ when transforming into the CDW2 phase. Note that the $T_{CDW}$ estimated from the Raman data at a given pressure is to certain extent lower than that from the transports and this effect becomes more evident at higher pressures, even after taking the laser heating effect into account, which may be related to combined effects of the sensitivity difference of these two experimental techniques in response to a CDW transition, the specific criterion we adopted in defining $T_{CDW}$ as well as a change of the CDW mechanisms (see Figures S8 and S9 and Supplemental Note 2).

The observation of emergence of a second CDW phase driven by pressure in an inherent CDW material is relatively scarce and to the best of knowledge, has only been verified in two CDW materials of $TiSe_2$[40] and $CsV_3Sb_5$[13-15]. In the former case a commensurate CDW transforms to an incommensurate one, while in the latter case the emergent CDW phase is ascribed to a possible stripe-like CDW order. The emergent CDW phase in both samples follows the same evolutionary trend of $T_{CDW}$. Unlike in these two materials, however, in CuTe its $T_{CDW}$ suddenly increases with accompanying the CDW transformation and $T_{CDW2}$ has a smaller decrease rate than that of $T_{CDW1}$, which suggests that these two CDW phases might be governed by different mechanisms.

We now discuss the intricated relationships between CDW and superconductivity for CuTe under high pressures. Accompanied by the suppression of the pristine CDW1 state, superconductivity is induced (Figure 3A). Along with the weakening of CDW1, $T_C^{onset}$ increases gradually and reaches a maximum of ~2.3 K. Subsequently, the CDW2 shows up and the $T_C^{onset}$ begins to decrease, resulting in a dome-like superconducting phase diagram. Meanwhile, the reduced superconducting transition $\Delta T_C/T_C^{onset}$ displays extraordinary broadening between 5 and 10 GPa and the recovery of a sharp superconducting transition is concomitant with the disappearance of the CDW2 phase (Figure 3B). As discussed above, a pressure-induced CDW transformation was also observed in $CsV_3Sb_5$, where an unexpected superconducting broadening with emergence of the new CDW. Moreover, an unusual M-shaped double superconducting dome was observed. These observations were considered as a signature of unusual competition between CDW and superconductivity in pressurized $CsV_3Sb_5$.[13-15] Such kind of phase diagram is in contrast to the present case of CuTe with one superconducting dome. Since the maximum $\Delta T_C/T_C^{onset}$ locates around the boundary of the CDW1→CDW2 transition, strong CDW fluctuations may be expected and lead to the observed superconducting broadening. Another possible mechanism for a superconducting broadening might be the CDW phase separation scenario. In pressurized TiSe$_2$,[40] optimal superconductivity is related to the CDW phase separation. Namely, a dome-like superconducting phase diagram is observed with the maximized $T_C$ achieved inside the domain wall of commensurate and incommensurate CDW phases. Nevertheless, no evident superconducting broadening is observed in that case,[11] thus seemingly not supporting the phase separation scenario as the possible mechanism of superconducting broadening in pressurized CuTe. Furthermore, since no evident indications of coexistence of CDW1 and CDW2 are observed from the transports and Raman data and relatively the pressure step is not small enough, some form of CDW1-CDW2 phase separation, if any, might be expected to appear only within a narrow pressure window of ~1-2 GPa. Note that sample/pressure inhomogeneity should not be as possible source of broadening since the situation would get worse upon further compression, which is at odds with our experimental observation of a sharp superconducting transition above ~10 GPa. Apparently, additional efforts are required to clarify this issue.

**Origin of the two distinct CDW orders**

In order to reveal the nature of the pressure-induced CDW transition, we calculated the orbital-resolved bands and Fermi surface, the static Lindhard response function and the phonon spectra with linewidth included for CuTe at various high pressures, which are shown in Figure 4 and Figures S10-S14. From Figures 4A and 4B (for more details, see Figures S10-S12), one can find that compression has two remarkable effects on the fermiology: (i) One band with green color near the Z-U line is pushed down at 8.3 GPa in comparison with that at ambient pressure. The consequence of effect (i) is pressure-induced effective electron doping, which can be explicitly found by comparing the band and Fermi surface near the Z-U line, and is consistent with the negative Hall coefficient $R_H$ shown in Figure S2 at pressures above 6.6 GPa. (ii) Four small isotropic hole-like pockets emerge when pressure is larger than ~6.0 GPa, which will be discussed below. Note that this pressure is close to the critical pressure of the CDW1-to-CDW2 transition.

Regarding the CDW ordering, there are typically three possible mechanisms involved in the CDW formation: (1) Peierls-like instability indicated by the nesting condition from the Lindhard response function, (2) strong momentum-dependent electron-phonon coupling and (3) electronic correlated interaction driven instability. We will evaluate each case to determine the most probable mechanism for the CDW formation in CuTe.

For case (1), the bare Lindhard response function $\chi(q,\omega)$ can be calculated by using the charge density-density correlation function, i.e.,

$$\chi(q,\omega) = \chi'(q,\omega) + i\chi''(q,\omega), \qquad \text{(Equation 1)}$$

with $\chi'(q,\omega)$, $\chi''(q,\omega)$ being the real and imaginary parts of the response function, respectively. $\chi'(q,\omega)$ describes the stability of the electronic system and $\chi''(q,\omega)$ evaluates the Fermi surface nesting. In the static situation $\omega = 0$ and at some $q = Q_c$, divergence for both $\chi'(Q_c, 0)$ and $\chi''(Q_c, 0)$ usually indicates the CDW phase transition. For unpressurized CuTe, such $Q_c$ exists and is equal to 0.4*2π/a, as shown in Figures S13A and S13B. This value is consistent with the wave vector where the imaginary frequency of phonon spectra emerges in Figure 4C. This can explain the CDW1 in unpressurized CuTe as driven by the nesting condition. However, no such $Q_c$ exists for CuTe at 8.3 GPa, as shown in Figures S13C and S13D. Additionally, the phonon spectra for CuTe at 8.3 GPa show no imaginary frequencies. This indicates that the CDW2 is not related to the nesting condition, which is different from that of CDW1.

For case (2), we calculated the phonon linewidth $\gamma_q$ the lowest acoustic phonon spectra, which indicates the strength of electron-phonon coupling. The results are shown in Figures 4C and 4D for CuTe at 0 and 8.3 GPa, respectively (also see Figure S14). Figure S14C gives exact values of $\gamma_q$ of the lowest acoustic phonon spectra for CuTe at 0 and 8.3 GPa, respectively. It can be observed that the values of $\gamma_q$ for the lowest acoustic phonon spectra in CuTe at 0 and 8.3 GPa are very small. We further evaluate the renormalized effect on the phonon spectra due to the electron-phonon coupling. Taking the high-symmetry $\Gamma - X$ line as an example, we know the following relations,

$$\gamma_q = 2|g(q)|^2 \chi''(q, 0), \qquad \text{(Equation 2)}$$
$$\omega(q)^2 = \omega_0(q)^2 - 2\omega_0(q)|g(q)|^2 \chi'(q, 0). \qquad \text{(Equation 3)}$$

Here, $g(q)$ is the electron-phonon coupling matrix element, $\omega_0(q)$ is the bare phonon spectra and $\omega(q)$ is the renormalized phonon spectra. The results are shown in Figure S14D. Compared with the bare phonon spectra labelled by the red line, the electron-phonon coupling renormalized phonon labelled by the blue line softens slightly but far from zero for CuTe at 8.3 GPa. It means that such weak electron-phonon coupling of the lowest acoustic phonon spectra here is not strong enough to drive the CDW2 instability in CuTe at 8.3 GPa.

For case (3), considering the failure of both cases (1) and (2), we may need to seriously consider the effect of electronic correlated interaction on CDW2. Figures 4A, 4B, 4E and 4F show the bands and the Fermi surface of the α band for CuTe at different pressures. It can be observed that there exist four small hole-like pockets with nearly full $p_z$ orbital weight in CuTe at 8.3 GPa. Unlike other patches of the Fermi surface with strong anisotropy, the nearly isotropy of the four small Fermi pockets forces the electron correlated interaction also to be isotropic. In such a case, the interaction driven CDW instability can be understood by the mechanism proposed by A. W. Overhauser,[41-43] i.e., the increase of unfavorable kinetic energy and static electric-field induced by charge density spatial modulation of the four small Fermi pockets can be balanced by the exchange and correlation energy from the electron correlated interaction and lattice distortions.

Altogether, we argue that case (3), i.e., the electronic correlated interaction, may be the most plausible mechanism of CDW2 in CuTe. In addition, the wave vector of CDW2 can be estimated as $Q_{c2} \sim (2k_F + Q_G)(1 + \frac{V_0}{4E_F})$,[43] where $k_F$ is the Fermi momentum and $Q_G$ connects two Fermi pockets similar to the umklapp processes. We expect $Q_{c2} \sim 0.6*2\pi/a(1+\delta)$ with $\delta \ll 1$ for the CDW2 phase. Note that $Q_{c2}$ is quite close to $Q_{c1}$, but has a different driving force.

**Conclusion**

We have performed a comprehensive high-pressure study on the evolution of CDW in CuTe. We demonstrated a pressure-induced superconductivity at ca. 4.8 GPa and a transition between two CDW orders at ca. 6.5 GPa. Our analyses and first-principles calculations provide evidence that the two CDW orders have different origins—CDW1 driven by the nesting condition from the Lindhard response function and CDW2 from electronic correlated interaction. Associated with the CDW transformation, the superconducting transition exhibits anomalous broadening. These observations of unusual evolutions of both superconductivity and CDW order in such a clean and simple binary compound CuTe demonstrate an ideal platform for studying the interplay between these two electronic orders.

EXPERIMENTAL PROCEDURES

**Resource availability**

*Lead contact*

Further information and requests for resources should be directed to and will be fulfilled by the lead contact, Xuliang Chen (xlchen@hmfl.ac.cn).

*Materials availability*

The materials generated in this study are available from the lead contact upon request.

*Data and code availability*

Requests for the data and code utilized in this work will be handled by the lead contact, Xuliang Chen (xlchen@hmfl.ac.cn).

**Material syntheses and high-pressure measurements**

CuTe single crystals were grown by a solid-state melting method.[31] High-pressure electrical transport experiments were performed by using a nonmagnetic Be-Cu alloy diamond anvil cell with NaCl powder as the pressure transmitting medium. The sub-kelvin environment of 400 mK was provided by a 9T C-MAG cryostat from CRYOMAGNETICS, Inc., with a He3 insert. In the variable temperature high-pressure Raman measurement, a freshly cleaved sample was loaded in a Be-Cu alloy diamond anvil cell made by DACTools (http://www.dactools.com/), with Daphne 7373 as the pressure-transmitting medium. The cell was placed in a JANIS-ST500 cryostat. Raman instruments were manufactured by Renishaw and the laser wavelength was 532 nm. The laser power used in the experiment was 2.5 mW. The laser heating effect was quantified by the ratio of the anti-Stokes to Stokes Raman mode intensity and we found it to be 15-25 K.[44] Note that the Raman spectra below 70 K were not collected since the background signal abruptly becomes extremely evident and even is comparable to the sample signal for unknown reasons. The Ruby fluorescence method was employed to determine the pressure at room temperature for all of the high-pressure experiments.

**First-principles calculations**

Based on the density functional theory (DFT), the first-principles calculations were carried out using the Vienna *ab initio* simulation package (VASP) program[45], with generalized gradient approximation (GGA) developed by the Perdew, Burke, and Ernzerhof functional (PBE)[46] and by using QUANTUM ESPRESSO (QE) package for lattice dynamics calculations, including phonon spectra and electron-phonon coupling[47]. The lattice parameters were obtained from Ref.[31] and all atoms were fully relaxed until the Hellmann-Feynman forces on each atom were smaller than 0.001 $eV/Å$ in structure relaxations. We adopted the PBE + U calculation method to deal with the $3d$ orbitals of Cu with $U_{eff} = 9\ eV$. The vdW corrections were included by the DFT-D3 method.[48] The phonon calculations were performed using the PHONOPY package.[49]

**SUPPLEMENTAL INFORMATION**

Supplemental Information can be found online at XXXX

**ACKNOWLEDGMENTS**

The authors are grateful for the financial support from the National Key R&D Program of China (Grants No. 2022YFA1602603 and No. 2022YFA1403200), the National Natural Science Foundation of China (Grants No. 12174395, No. U19A2093, No. 12004004, No. 12204004, No. 12022413, No. 92265104, and No. 11674331), the Users with Excellence Program of Hefei Center CAS (Grant No. 2021HSC-UE008), the Collaborative Innovation Program of Hefei Science Center CAS (Grants No. 2020HSC-CIP014 and No. 2020HSC-CIP002), the "Strategic Priority Research Program (B)" of the Chinese Academy of Sciences (Grant No. XDB33030100), and the Major Basic Program of Natural Science Foundation of Shandong Province (Grant No. ZR2021ZD01). Y.H.Z. was supported by the Youth Innovation Promotion Association CAS (Grant No. 2020443).

**AUTHOR CONTRIBUTIONS**

X.L.C. and Z.R.Y. conceived the original idea and supervised the project. S.Y.W. grew the single crystals. S.Y.W., C.A. and Y.H.Z. performed the high-pressure electrical transport measurements. S.Y.W., C.A. and Y.Z. performed the high-pressure Raman experiments. N.H. and Q.W. performed the DFT calculations. Z.R.Y., N.H., X.L.C., and S.Y.W. analyzed the data and wrote the manuscript with contributions from all authors.

## DECLARATION OF INTERESTS

The authors declare no conflict of interest.

## INCLUSION AND DIVERSITY

We support inclusive, diverse, and equitable conduct of research.

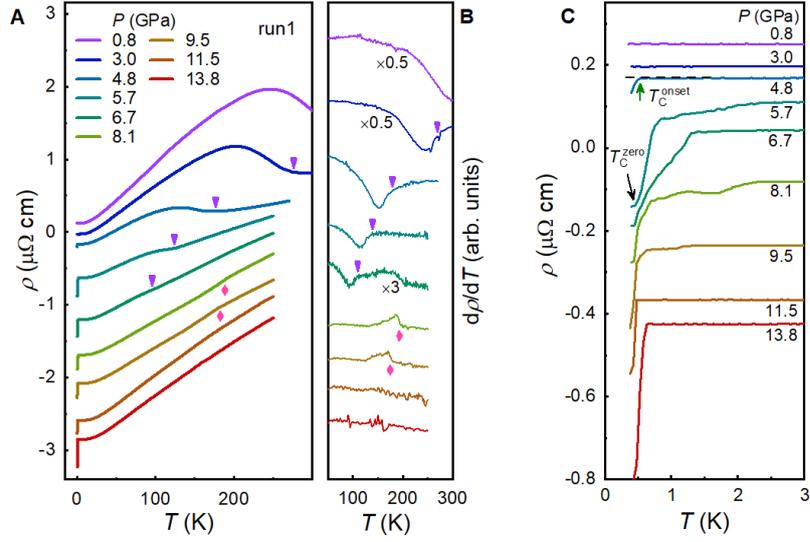

**Figure 1. Temperature dependence of resistivity of single-crystalline CuTe under high pressure**

(A and B) Resistivity ($\rho$) (A) and its derivative d$\rho$/d$T$ (B) for CuTe in the pressure range 0.8–13.8 GPa. The transition temperatures where hump-like and kink-like anomalies onset are marked by the purple triangles and pink rhombuses, respectively. All the curves have been shifted vertically for clarity.

(C) The evolution of the superconducting transition at various pressures. The green and black arrows indicate $T_C^{onset}$ and $T_C^{zero}$, respectively. All the curves have been shifted vertically for clarity.

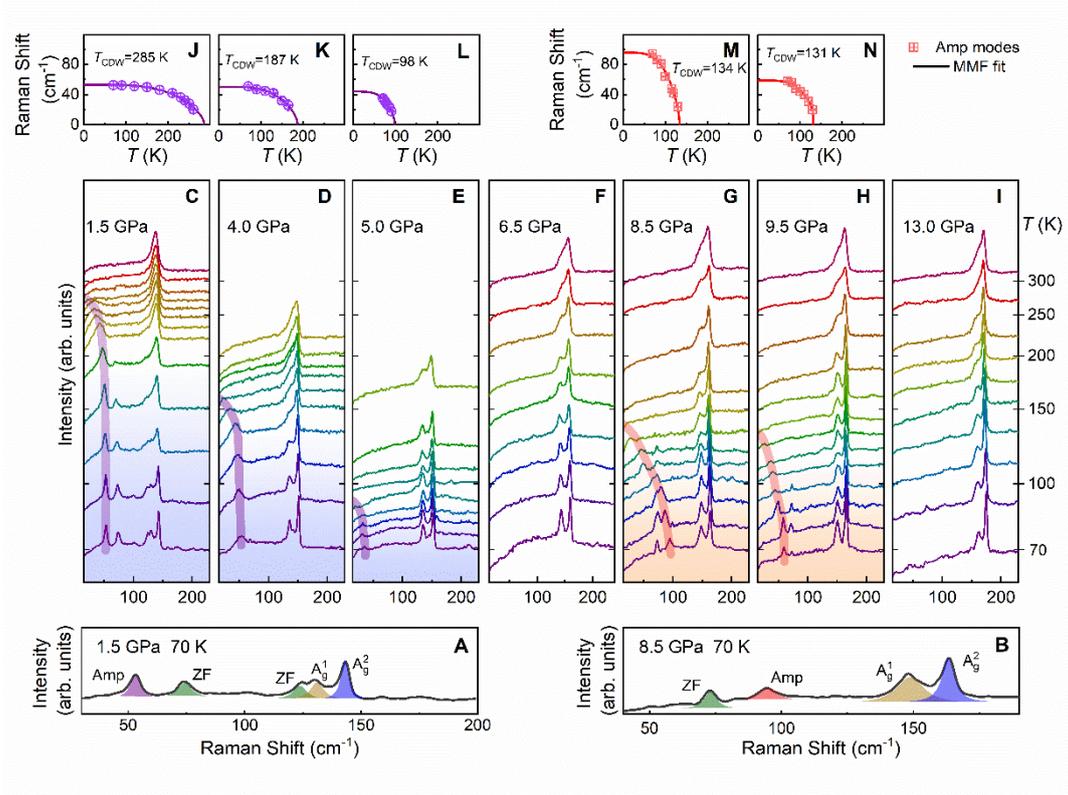

**Figure 2. Low-temperature Raman spectra of CuTe at high pressures**

(A and B) Raman spectra of CuTe at 70 K under 1.5 GPa and 8.5 GPa, respectively. Amp denotes the CDW amplitude mode.

(C-I) Temperature-dependent Raman spectra at various pressures. The lines are guides to the eyes highlighting the evolution of the amplitude mode with temperature.

(J-N) Raman shift of amplitude mode as a function of temperature at different pressures. The solid lines represent fits using the modified mean-field (MMF) model.

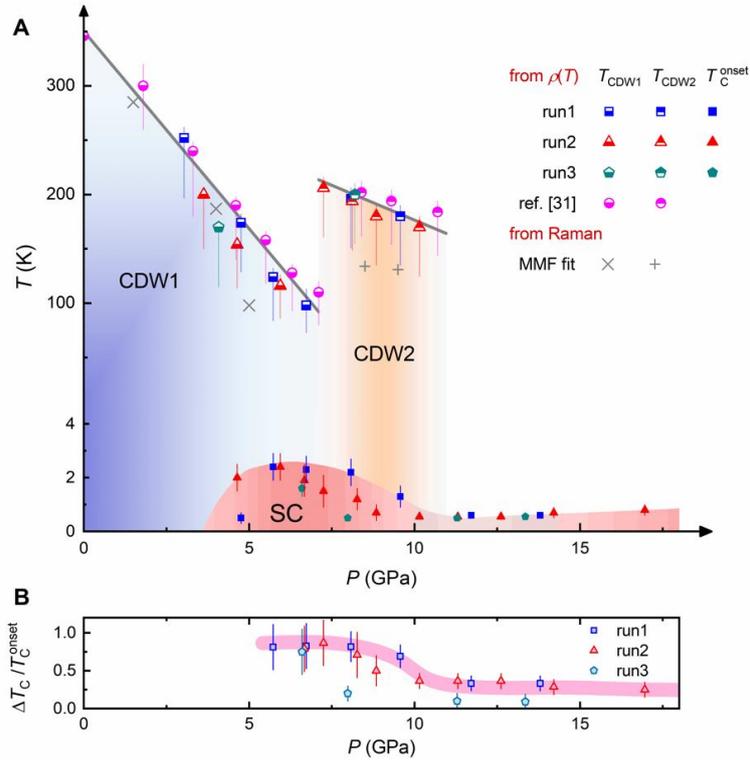

**Figure 3. Temperature-pressure phase diagrams of CuTe**

(A) Pressure dependences of the CDW transition temperatures $T_{CDW1}$ and $T_{CDW2}$ and the superconducting transition temperature $T_C^{onset}$.

(B) The reduced superconducting transition width $\Delta T_C/T_C^{onset}$ as a function of pressure. The red line is a guide to the eyes.

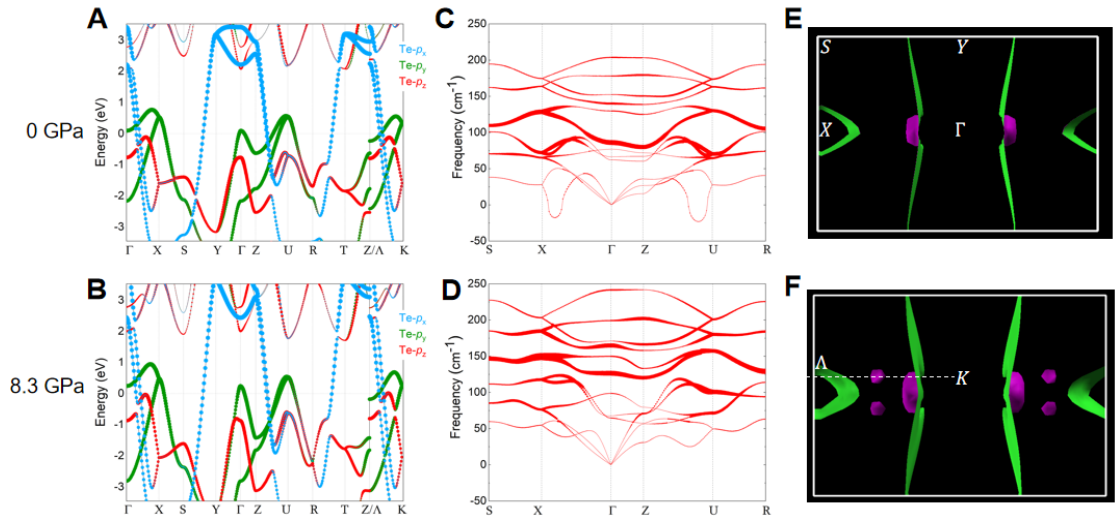

**Figure 4. Calculated electronic band structures and phonon spectra**

(A and B) The band structures of CuTe along high-symmetry lines in the non-CDW states at 0 and 8.3 GPa, respectively. The band structures of Te $p_x$, $p_y$ and $p_z$ are labelled with blue, green and red, respectively.

(C and D) The phonon spectra along high symmetry lines at 0 and 8.3 GPa, respectively. The phonon line width is indicated by the thickness of the spectra.

(E and F) Top view of Fermi surface sheets from α band at 0 and 8.3 GPa, respectively.